\documentclass[a4paper,11pt]{article}
\pdfoutput=1 

\usepackage{jheppub} 
\usepackage{amsmath}
\usepackage{pict2e}
\usepackage{physics}
\usepackage{amsfonts}
\usepackage{mathtools}
\usepackage{amssymb,bbm}
\usepackage{dsfont}
\usepackage{subcaption}
\usepackage{braket}
\usepackage{amsthm}
\usepackage{xcolor}
\usepackage{tikz}
\usepackage{ytableau}
\usepackage{graphicx}
\usetikzlibrary{arrows.meta}
\usepackage{hyperref}

\makeatletter
\newcommand{\farsquare}[2]{#1\,{\mathpalette\far@square{#2}}}
\newcommand{\far@square}[2]{%
  \mathop{\vcenter{\hbox{%
    \sbox\z@{$\m@th#1\sum$}%
    \setlength{\unitlength}{0.9\dimexpr\ht\z@+\dp\z@}%
    \begin{picture}(1,1)
    \roundjoin
    \polyline(0,0)(0,1)(1,1)(1,0)(0,0)(0,0.5)
    \end{picture}%
  }}}\limits_{#1#2}%
}
\makeatother

\newcommand{\nc}{\newcommand}
\nc{\rnc}{\renewcommand} 

\rnc{\a}{\alpha}
\rnc{\b}{\beta}
\nc{\g}{\gamma}
\rnc{\d}{\delta}
\nc{\e}{\epsilon}
\nc{\ee}{\varepsilon}
\nc{\z}{\zeta}
\nc{\f}{\phi}
\nc{\m}{\mu}
\nc{\n}{\nu}
\rnc{\r}{\rho}
\rnc{\k}{\kappa}
\rnc{\l}{\lambda}
\nc{\p}{\pi}
\nc{\s}{\sigma}
\rnc{\t}{\tau}
\nc{\w}{\omega}
\nc{\x}{\chi}
\nc{\F}{\Phi}
\rnc{\L}{\Lambda}

\nc{\1}{\mathbbm{1}}

\usepackage[draft,inline,nomargin,index]{fixme}
\FXRegisterAuthor{kr}{akr}{\color{blue}KR}

\title{\boldmath A note on entanglement entropy and  topological defects  in symmetric  orbifold CFTs}

\author[a]{Michael Gutperle,}
\author[a]{Yan-Yan Li,}
\author[a]{Dikshant Rathore,}
\author[b]{and Konstantinos Roumpedakis}

\affiliation[a]{
Mani L. Bhaumik Institute for Theoretical Physics, Department of Physics and Astronomy,\\ University of California, Los Angeles, CA 90095, USA.
}
\affiliation[b]{William H. Miller III Department of Physics and Astronomy, Johns Hopkins University,\\ 3400 North
Charles Street, Baltimore, MD 21218, USA.
}

\emailAdd{gutperle@ucla.edu}
\emailAdd{yanyanli@ucla.edu}
\emailAdd{drathore@physics.ucla.edu}
\emailAdd{kroumpe1@jh.edu}

\abstract{In this brief note we calculate the entanglement entropy in $M^{\otimes N}/S_N$ symmetric orbifold CFTs in the presence of topological defects, which were recently constructed in \cite{Gutperle:2024vyp,Knighton:2024noc}. We consider both universal defects which realize $Rep(S_N)$ non-invertible symmetry and non-universal defects.  We calculate the sub-leading defect entropy/g-factor for defects at the boundary of the entangling surface as well as inside it.}

\begin{document} 
\maketitle
\flushbottom

\section{Introduction}
\label{sec:intro}

Entanglement is one of the most important features of quantum theories and a specific measure of entanglement, namely entanglement entropy, has played an important role in many areas of physics ranging from quantum mechanics, condensed matter physics, quantum information theory to quantum field theory and quantum gravity, holography and the black hole information paradox, see \cite{Calabrese:2009qy,Nishioka:2018khk,Casini:2022rlv} for a selection of reviews of some of these topics.  

Topological defects in two dimensional conformal field theories (CFTs)  \cite{Verlinde:1988sn,Petkova:2000ip,Petkova:2001ag,Fuchs:2002cm,Chang:2018iay} are among the best understood examples of non-invertible symmetries, displaying a fusion algebra structure, defect lines ending on defect operators, junctions of defect lines, etc. More recent ideas are the constraints on RG-flows from non-invertible symmetries, symmetry topological field theories, and half gauging among others\footnote{See \cite{Shao:2023gho,Bhardwaj:2023kri,Schafer-Nameki:2023jdn} recent lecture notes for more details and a comprehensive list of references.}.

In this short note, we calculate the entanglement entropy for topological defects in symmetric orbifold CFTs which were recently constructed in \cite{Gutperle:2024vyp,Knighton:2024noc}, using the methods developed for topological defects in rational CFTs in \cite{Sakai:2008tt,Brehm:2015plf,Gutperle:2015kmw}. The defect entropy we calculate can be viewed as a measure of the complexity of the defect. Furthermore, it provides an observable which may be of use in calculating the entanglement entropy holographically using the dual   $AdS_3\times S^4\times \mathcal{M}_4$ background with the minimal $k=1$ unit of NS-fivebrane flux \cite{Gaberdiel:2017oqg,Gaberdiel:2018rqv,Eberhardt:2018ouy,Giribet:2018ada,Eberhardt:2019ywk}. Such a calculation may provide a further check on the proposals of the dual of the topological defect operators put forward in \cite{Gutperle:2024vyp,Knighton:2024noc}.

We review the formalism to calculate entanglement entropy in the presence of (topological) defects in section \ref{sec:ee1} and the construction of topological defects in symmetric orbifold CFTs in section \ref{sec:orb}. We calculate the entanglement entropy both for universal and non-universal defects in section \ref{sec:ee2}. We close the note with a brief discussion of some future research directions in section \ref{sec:Discussion}.

\section{Engtanglement entropy and topological defects}
\label{sec:ee1}
In two dimensional CFTs, the simplest example of entanglement entropy is given by considering the theory in its vacuum and a single spatial interval ${\cal A}$. The entanglement entropy is then given by the von Neumann entropy of the reduced density matrix $\rho_{\cal A}= \tr_{\cal \bar A} | 0\rangle \langle 0|$, where we trace over the states in the complement of ${\cal A}$.
\begin{align}
S_{\cal A} = -\tr \rho_{\cal A} \ln \rho_{\cal A}~.
\end{align}
The replica trick  allows the entanglement entropy to be calculated from  R\'enyi entropies 
\begin{align}
S_{\cal A}  =  -{\partial_K} \tr \rho_{\cal A}^K  \Big |_{K=1}~.
\end{align}
The $K$-th R\'enyi entropy can be calculated by a path integral over a  Riemann surface with $K$ sheets, with branch  cuts along the interval ${\cal A}$ connecting the sheets in a cyclic fashion. The entanglement entropy for the CFT in its ground state can be obtained from the partition function $Z(K)$ over the $K$-sheeted Riemann surface, by the following expression
\begin{align}\label{entee1}
S_{\cal A}=\big(1- \partial_K\big) \ln Z(K) \Big|_{K=1}~.
\end{align}
In the evaluation of the path integral one  has to introduce an ultraviolet cutoff $\epsilon$ and the terms which survive the limit of the cutoff going to zero are
\begin{align}\label{ee-general}
S_{\cal A}=\frac{c}{ 3} \ln  \frac{L}{ \epsilon } +\ln g_{\cal A}~,
    \end{align}
where $c$ is the central charge of the CFT. The constant term is not universal in the case of the CFT defined on a real line in the ground state. However, it can become physical in the the presence of a boundary or interface \cite{Cardy:2004hm, Azeyanagi:2007qj,Chiodaroli:2010ur} and is sometimes called the g-factor or  boundary entropy \cite{Affleck:1991tk}. In the following we would like to calculate this g-factor for topological defects in symmetric orbifold CFTs.

\subsection{Entanglement of symmetric defect}

The g-factor \eqref{ee-general} also appears in  boundary  CFTs,  where the partition function of a CFT is defined on an interval of length $L$, with boundary conditions labeled by $a$ and $b$ at the two ends of the interval.
\begin{align}
    \ln Z = \ln \tr e^{-\beta H_{ab}}  \sim \ln( g_a g_b) + \frac{c\pi}{6} \frac{L}{\beta} +\cdots,
\end{align}
which is valid in the limit $L>> \beta$. After a modular transformation from the open string annulus, the partition function can be viewed as  a closed string cylinder between two boundary states 
\begin{align}
Z=\langle \langle a| e^{-L H_{cl}}|b\rangle\rangle~.
\end{align}
One can extract the g-factors by inserting the ground state and taking the $L\to \infty $ limit. It follows that
\begin{align}\label{gfac-bcft}
    g_a= \langle 0| a\rangle\rangle~.
\end{align}

It has been shown \cite{Azeyanagi:2007qj} that for an entangling surface $\mathcal{A}$  which starts at the boundary, the g-factor appearing in the entanglement entropy   (\ref{ee-general}) is given by (\ref{gfac-bcft}).
So far we have considered a boundary CFT. For a conformal interface, one can use the folding trick
\cite{Oshikawa:1996dj,Bachas:2001vj} where an interface between CFT$_1$ and CFT$_2$ is turned into a boundary CFT by folding CFT$_1\times \overline{\text{CFT}}_2$, which can be represented by a boundary state $| B\rangle\rangle$. It was argued in  \cite{Azeyanagi:2007qj,Chiodaroli:2010ur} that the g-factor for the entanglement entropy for a conformal defect which is symmetric (i.e. in the middle of the entangling region ${\cal A}$) is  given by (\ref{gfac-bcft}), where $\mid a\rangle \rangle$ is the boundary state representing the defect in the folded CFT.
\subsection{Entanglement entropy of defects at the entangling surface }

\begin{figure}[h]
\centering
\includegraphics[scale=.39]{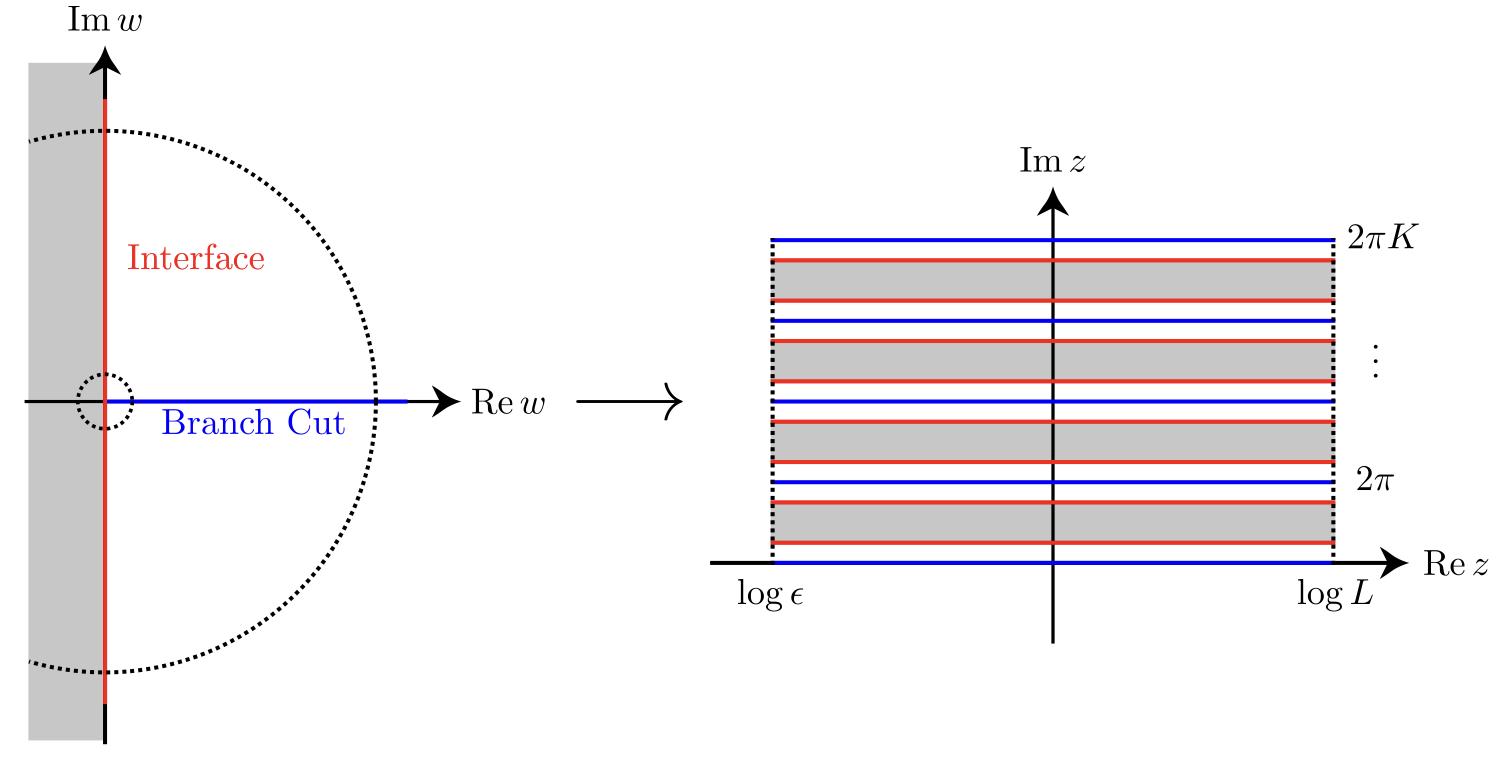}
\caption{Conformal map for the replicated worldsheet for defects at the entangling surface, figure from \cite{Gutperle:2015kmw}.}
\label{Fig1}
\end{figure}

In order to describe the defect located at the boundary of the entangling region ${\cal A}$ and its complement (see the left-hand side of Figure \ref{Fig1}), we can introduce the $K$ sheeted covering map with $z=\ln w.$
We introduce a UV cutoff $\epsilon$ at the boundary of ${\cal A}$ and an IR cutoff $L$ as the length of the branch cut. In general, we have to impose boundary conditions at the cutoff surfaces with the simplest one being to identify the cutoff surfaces periodically, which produces a toroidal worldsheet\footnote{More general conformal boundary conditions have been considered in constructions of symmetry resolved entanglement entropy in  \cite{Cardy:2004hm,Ohmori:2014eia, DiGiulio:2022jjd, Kusuki:2023bsp}.}. It is possible to place a conformal interface at the boundary, which in the replicated surface will turn into $K$ copies of ${\cal I}$ and $K$ copies of the conjugate defect ${\cal I}^\dagger$ placed equidistant on the torus
\begin{align}\label{zreplica}
Z(K) ={\rm tr} \Big[ \big( {\cal I} e^{- t H } {\cal I}^\dagger e^{- t H }\big)^K \Big]~.
\end{align}
The Euclidean time $t$ can be expressed as
\begin{align}
t= \frac{2\pi^2 }{ \ln L/\epsilon}~,
\end{align}
 the simplest choice $\epsilon=1/L$ is made in \cite{Sakai:2008tt} and the limit $\epsilon\to 0$ corresponds to removing the UV cutoff as well as taking $L$ to be large. 
For conformal defects, the replica partition function has only been calculated for simple cases such as free bosons and fermions as well as minimal models \cite{Sakai:2008tt,Satoh:2011an,Brehm:2015lja,Gutperle:2015hcv}. In all examples, the entanglement entropy behaves as follows\footnote{The $c_{eff}/6$ differs from \eqref{ee-general} by a factor of $1/2$, this is caused by taking the limit $L\to \infty$ and only taking the entanglement of one side of the entangling surface into account, see \cite{Karch:2021qhd} for a discussion on this issue.}
\begin{align}\label{ee-interv}
    S_{\cal A} =\frac{c_{\rm eff}} { 6} \ln \frac{L}{ \epsilon}+ \ln g_{\cal A}~,
\end{align}
where in general $c_{\rm eff}$ is not the central charge $c$ of the bulk CFT but instead depends, in a complicated fashion, on the details of the conformal interface (see \cite{Karch:2024udk} for a recent discussion on universal properties of $c_{\rm eff}$).

For topological defects, the expression  (\ref{zreplica}) simplifies since the defects commute with the Hamiltonian and they can be combined
\begin{align}\label{zreplica2}
Z(K) ={\rm tr} \Big[ \big( {\cal I}{\cal I}^\dagger\big)^K  e^{- 2 t K H } \Big]~.
\end{align}
The entanglement entropy derived from this partition function has the form (\ref{ee-interv}) with $c_{\rm eff}=c$. Hence, the leading term in the entanglement entropy does not depend on the defect and all information is in the sub-leading g-factor which contains information about the ground state degeneracy of the interface. It is the goal of the present paper to calculate this quantity for the topological defects in symmetric orbifolds constructed in  \cite{Gutperle:2024vyp}.

\section{Topological defects in symmetric orbifolds}
\label{sec:orb}
In \cite{Gutperle:2024vyp} the present authors constructed topological defects in symmetric orbifold CFTs. In this section, we will briefly review the main results of that paper and refer the reader to it for more details.

The  ${\cal M}^N/S_N$ symmetric orbifold is constructed from a seed CFT ${\cal M}$ by taking the $N$-fold tensor product and gauging/orbifolding the $S_N$ permutation symmetry.
There are two kinds of defects that have been constructed: first, the so-called universal defects, which do not depend on the details of the seed theory such as the spectrum, fusion coefficients, etc. In  \cite{Gutperle:2024vyp} (see also \cite{Knighton:2024noc}), in addition to the universal defects, one kind of non-universal defect was constructed, which is built on a single nontrivial topological defect in the seed CFT  and was called ``maximally fractional"  following a similar construction for boundary CFTs in symmetric orbifolds as in \cite{Gaberdiel:2021kkp}. In the following, we will briefly review both constructions.

\subsection{Symmetric orbifold}

An $S_N$ symmetric orbifold CFT is constructed by taking $N$ copies of the seed CFT ${\cal M}$ and gauging the permutation symmetry. The untwisted sector is obtained by projecting out all operators in ${\cal M}^{\otimes N }$ which are not invariant under $S_N$. Examples of gauge invariant operators are
\begin{equation}
    \sum_{I=1}^N O_I, \quad  \sum_{I,J=1}^N O_IO_J, \quad \dots ~,
\end{equation}
where $I$ labels the copy of the seed CFT in the tensor product.
Modular invariance enforces the presence of twisted sector fields $\sigma_{g}$, where $g\in S_N$, which acts on an operator in the $I$-th copy as a permutation when moved around a twist field
\begin{equation}
    O_I(e^{2\pi i }z) \; \s_g(0) = O_{g\cdot I}(z)\; \s_g(0)~. \label{eq:twist_def}
\end{equation}
To define gauge invariant operators, we sum over all conjugations by elements of $S_N$  \cite{Dijkgraaf:1989hb, Lunin:2000yv, Pakman:2009zz}
\begin{equation}\label{eq:g-inv_def}
    \s_{[g]} \propto \sum_{h \in S_N}  \s_{h^{-1} g h}~,
\end{equation}
which are labeled by the conjugacy classes $[g]$ of $S_N$.  These classes are completely determined by giving the number $l_i$ of cycles of length $i=1,2,\cdots N$ in a permutation.
\begin{align}\label{cyclemg}
    [g] = { 1} ^{l_1} \cdot  { 2}^{l_2} \cdots {N}^{l_N}~,
\end{align} where the $l_k$ are a partition of $N$, i.e. they satisfy $\sum k \; l_k=N$.
The torus  partition function can be written as follows \cite{Klemm:1990df}
\begin{align}\label{symob-z}
Z &= \frac{1}{ |G|} \sum_{gh=hg}   \farsquare{h}{g} \; (\tau,\bar \tau)~,
\end{align}
 where $|G|=N!$ is the order of $S_N$.
Here $g$ denotes the twisted sector and $h$ represents elements of its centralizer.  Each term in the sum is equal to  
\begin{align}\label{zhg-part}
 \farsquare{h}{g} \; (\tau,\bar \tau) = {\rm tr}_{{\cal H}_g} \Big( h e^{2\pi i \tau (L_0-\frac{c}{ 24}) } e^{-2\pi i \bar \tau (\bar L_0-\frac{c}{ 24}) } \Big)~.
\end{align}
The summation of elements $h\in S_N$ which commute with $g$ imposes the projection onto states that are invariant under the centralizer $C[g]$ and ensures the gauge invariance of all the states. 
Using the fact that the torus partition function in the individual sectors transforms as follows under the modular transformations $\tau\to -\frac{1}{ \tau}$ 
 \begin{align}\label{modular-a1}
  \farsquare{h}{g} (\tau, \bar \tau) \to  \farsquare{g^{-1}}{h} (-\frac{1}{\tau},-\frac{1}{  \bar \tau}) ~,
 \end{align}
it is straightforward to check that the orbifold partition function  (\ref{symob-z}) is indeed modular invariant. 

\subsection{Universal defects}
\label{sec:universald}

Here, we employ the folded permutation brane construction for topological defects \cite{Recknagel:2002qq,Drukker:2010jp,Cordova:2023qei}. Utilizing this construction,  a universal defect corresponds to the following boundary state in the folded $M^N/S_N \times M^N/S_N$ CFT, where states in the two factors are labeled by subscripts $i=(1),(2)$.
 \begin{align}\label{boundary-state}
 \mid B_R\rangle \rangle&= \sum_{[g]}  \chi_R([g])  \sum_n  \mid n\rangle_{(1),[g]} \otimes \overline{ \mid n\rangle}_{(2),[g]} \sum_m  \mid m\rangle_{(2),[g]} \otimes \overline{ \mid m\rangle}_{(1),[g]} ~,
 \end{align}
here $[g]$ denotes the twisted sector in the $M^N/S_N$ orbifold labeled by the conjugacy class $[g]$. The summation over $n,m$ denotes the sum over all states in the $[g]$-twisted sector including the sum over primaries of the seed theory as well as (twisted) Virasoro descendants and produces an Ishibashi-like boundary state \cite{Ishibashi:1988kg,Cardy:1989ir}.
The folded boundary state satisfies
 \begin{align}\label{boundary-a}
     \Big(L^{(1)} _n-\bar L^{(2)}_{-n}\Big) \mid B_R\rangle\rangle&=  \Big(L^{(2)} _n-\bar L^{(1)}_{-n}\Big) \mid B_R\rangle\rangle =0~,
 \end{align}
where $L^{(i)}_n, i=1,2$ are the modes of the stress tensor of the two copies of the $M^N/S_N$ orbifold after folding the defect. After unfolding the two copies of the CFT living on half-spaces, one gets a single copy of the $M^N/S_N$ orbifold  CFT with a defect operator 
 \begin{align}\label{td-projector}
{\cal I} _R&=  \sum_{[g]}  \chi_R([g]) {P}_{[g]}\bar { P}_{[g]} ~.
\end{align} 
Here, ${P}_{[g]}$ is the identity projector in the $[g]$-twisted sector.
It is clear from the property of the projectors or the unfolding of the boundary state condition (\ref{boundary-a}) that the defect is topological, i.e. it commutes with the holomorphic and anti-holomorphic modes of the stress tensor separately.
 \begin{align}\label{td-state}
     L_{n} {\cal I} _R  &= {\cal I} _R L_n, \quad  \quad \bar L_{n} {\cal I} _R  = {\cal I} _R \bar L_n~.
     \end{align}

\subsection{Non-universal defects}
\label{subsec:nonuni} 
The universal defects discussed in the previous section are written as projectors. They are proportional to the identity operator in each twisted sector. This implies that they do not depend on the details of the seed CFT, or put it in another way, they are built upon the trivial or identity defect in the seed CFT.

Here we will review the construction \cite{Gutperle:2024vyp}  of an example of a non-universal defect which is analogous to the ``maximally fractional" D-brane of \cite{Gaberdiel:2021kkp}\footnote{See also  \cite{Knighton:2024noc} for a more detailed discussion of these defects, in particular for the trivial and alternating representation of $S_N$.}. A twisted sector  (\ref{cyclemg}) in the $S_N$ orbifold theory is labeled by a conjugacy class $[g]$ which is fixed by giving the number $l_i$ of cycles of given length $i=1,2,\cdots N$.
 In addition, we will use a diagonal rational conformal field theory (RCFT) as the seed CFT. The partition function is given by
 \begin{align}
    Z(\tau,\bar{\tau}) &= \sum_{i,j} \delta_{ij} \;  \chi_i(\tau) \bar{\chi}_j(\bar{\tau})~,
\end{align}

where $i=1,2,\cdots, n$ labels the finite irreducible chiral vertex algebra with character $\chi_i(\tau)$. Under the modular $S$-transformation, the characters transform as $\chi_i(-1/\tau) = \sum_j S_{ij} \; \chi_j(\tau)$, where $S_{ij}$ is the modular S-matrix.  
The topological defects we will consider are  Verlinde lines ${\cal I}_a$ \cite{Verlinde:1988sn}, which are labeled by  $a$ denoting the primaries and can be constructed in terms of projectors
\cite{Petkova:2000ip,Petkova:2001ag}
\begin{align}\label{rcft defect}
    {\cal I}_a = \sum_i \frac{S_{ai}}{S_{0i}} P_{i \bar{i}}~.
\end{align}

In the tensor product states, we will use the same $a$ for all factors to construct a ``maximally fractional" 
 defect, which is automatically $S_N$  invariant.\footnote{For a more general construction 
 of D-branes in symmetric orbifolds, see \cite{Belin:2021nck}.}  In a twisted sector corresponding to a single $w$ cycle, the doubled boundary state is given by
\begin{align}\label{doubled-b}
\mid B_a\rangle \rangle_{(w)} = \sum_ i \frac{S_{ai}}{ S_{0i} }   \sum_{n} \ket{i, n}^{(1)}_{(w)} \otimes \overline{\ket{i,n}}^{(2)}_{(w)}   \sum_{m}   \ket{i,m}^{(2)}_{(w)} \otimes \overline{\ket{i,m}}^{(1)}_{(w)}~,
\end{align}
where the sum of $n,m$ denotes the sum over all the descendants (including fractional Virasoro modes) of the $w$-twisted sector primary $\ket{i}_{(w)}$. Using these boundary states for the cycles, we can construct a defect boundary state which is labeled by the representation $R$ of $S_N$ as well as the choice of primary $a$ of the RCFT.
\begin{align}\label{doubled-c}
\mid B_a^R\rangle \rangle&= \sum_{[g]\in S_N}  \chi_R([g]) \prod_{j=1}^N \prod_{k_j=1}^{l_j} \mid B_a\rangle \rangle_{(j)}~,
\end{align}
where the product goes over all conjugacy classes of $S_N$ labeled as in \eqref{cyclemg}.
In \cite{Gutperle:2024vyp} it was shown that these boundary states satisfy the Cardy conditions and therefore also define consistent topological defects in the sense of \cite{Petkova:2000ip,Petkova:2001ag} after unfolding the boundary state (\ref{doubled-c}).

\section{Entanglement entropy with symmetric orbifold defect}
\label{sec:ee2}

In this section, we will use the formulas reviewed in section \ref{sec:ee1} to calculate the entanglement entropy in the presence of the topological defects whose construction was reviewed in section \ref{sec:orb}.

\subsection{Universal defects}
\label{subsec4.1}
Recall that the universal defects constructed in section \ref{sec:universald} are labeled by the irreducible representation $R$ of $S_N$. Using the folded boundary state (\ref{boundary-state}), it follows from (\ref{gfac-bcft}) that the sub-leading contribution to the entanglement entropy for a symmetrically placed topological defect is given by the logarithm of the 
g-factor, which in turn is calculated by the overlap of the boundary state with the vacuum
\begin{align}\label{gfac-sym}
    g_{\cal A}= \langle 0 \mid B_R\rangle\rangle  =  \chi_R(1) = \dim(R)~.
    \end{align}
    Note that $g_{\cal A}$ is given by the quantum dimension of the defect, which is simply the dimension of the irreducible representation $R$. For conformal defects, this result is only valid for the symmetrically placed defect. For topological defects, we can move them away from the middle and the result will not change unless we come very close to the boundary of the entangling region.

Note, however, that this is not true for the topological defect at the boundary of the entangling surface as discussed in section \ref{sec:ee1}. For the universal topological defects, one finds
\begin{align}
Z(K) &= {\rm tr} \Big[  {\cal I}^{2K} e^{- 2 K t H}\big] \nonumber \\
&= \sum_{[g]} \big[\chi_R([g]) \big]^{2K} \tr\big( P_{[g]} \bar P_{[g]}  e^{- 2 K t H}\big)~,
\end{align}
where we used the reality of the characters which implies that ${\cal I }^\dagger ={\cal I}$.   The modular parameter is given by 
\begin{align}
\tau =  i \frac{2\pi K}{   \ln \frac{L}{ \epsilon}}~.
\end{align}
The partition function can be expressed as a sum over twisted sectors using the notation introduced in (\ref{zhg-part})
\begin{align}
Z(K) &= \frac{1}{|G|} \sum_{hg=gh} \big[\chi_R([g]) \big]^{2K}   \farsquare{h}{g}( \tau, \bar \tau)~.
\end{align}
We can view this as a summation over twisted sectors labeled by $g$ and a projection on $S_N$ invariant states implemented by $h$, where $h$ runs over the centralizer of $g$, i.e. all elements which commute with $g$.
  Taking the cutoff $L\to \infty$ implies that $\tau \to 0$, so we have to perform a modular transformation of the torus partition function

\begin{align} 
Z(K) &= \frac {1}{ |G|} \sum_g \sum_{h\in C[g]} \big[\chi_R([g]) \big]^{2K}  \farsquare{g^{-1} }{h }( - \frac{1}{ \tau} ,-\frac{1}{ \bar \tau})~.
\end{align}
Note that the limit $\tau \to 0$ implies that the argument of the modular transformed partition function behaves as  $-\frac{1}{ \tau }\to \infty$. Hence, the partition function is dominated by the sector with the primary of the lowest conformal dimension. For the symmetric orbifold, this is the untwisted vacuum sector.  

Note that in the centralizer $C[g]$ of any element $g\in S_N,$ the identity is always present since it trivially commutes with any element. Hence, the partition function in the $\tau\to 0$ limit is dominated by the untwisted sector
\begin{align}
\lim_{\tau \to 0}  Z(K) 
&\sim\lim_{\tau \to 0}   \sum_{g\in S_N} \frac{1}{|G|}  \big[\chi_R([g]) \big]^{2K} \farsquare{g }{1 }( -\frac{1}{ \tau} ,-\frac{1}{ \bar \tau}) ~,
\end{align}
where we also used the reality of the character to replace $g^{-1}$ by $g$ in the torus partition function.
In the limit $\tau \to 0$ each term in the partition function contributes the same leading term, which is independent of $g.$
\begin{align}\label{limita}
 \lim_{\tau\to 0} \farsquare{g }{1 } ( -\frac{1}{ \tau} ,-\frac{1}{ \bar \tau})  &=  e^{ - \frac{2\pi i}{  \tau}  (-\frac{c}{ 24})  }e^{ \frac{2\pi i}{  \bar \tau}  (-\frac{c}{ 24})  } + \cdots \nonumber\\ 
 &= \exp\Big( {\frac{c}{ 12 }  \frac{1}{ K} \ln \frac{L}{ \epsilon} }\Big)  \Big\{ 1 +  {\cal O}( e^{-  \frac{1}{ K}  \ln \frac{L}{ \epsilon} } ) \Big\}~,
 \end{align}
 where the central charge is $c= N c_{seed}$. 
 
 The argument for this is as follows: the modular invariant partition function of the seed theory is $Z(\tau, \bar \tau)$.  Consider an element $g\in S_N$ which is given by $k$ cycles of lengths $n_1,n_2, \cdots n_k$ with $ \sum_k n_k=N$. In the tensor product theory, the partition function with $g$ inserted is
\begin{align}
\farsquare{g }{1 }  (-\frac{1}{\tau}, -\frac{1}{ \bar \tau} ) = \prod_{i=1}^k  Z_{seed}(-  n_i \frac{1}{ \tau}  , -n_i \frac{1}{ \bar\tau})~.
\end{align}
In the limit $\tau \to 0 $  the partition function of the seed theory is dominated by the vacuum contribution 

\begin{align}
\lim_{\tau \to 0} \farsquare{g }{1 }  (-\frac{1}{ \tau}, -\frac{1}{ \bar \tau} ) &\sim  \prod_{i=1}^k  e^{- \frac{2\pi i }{ \tau }  n_i c_{seed}/24 } e^{\frac{2\pi i }{ \bar  \tau }  n_i c_{seed}/24 } \nonumber \\
&\sim e^{-\frac{2\pi i}{  \tau }  N c_{seed}/24 } e^{\frac{2\pi i}{ \bar  \tau }  N c_{seed}/24 } \nonumber \\
&\sim   \exp\Big( {\frac{c}{ 12 } \frac {1}{ K} \ln \frac{L}{ \epsilon} }\Big) ~,
\end{align}
where we used $\sum_i n_i =N$.
Putting these ingredients together  we obtain 
 \begin{align}
  \lim_{\tau \to 0}  \ln  Z(K)&=  {\frac{c}{ 12 }  \frac{1}{ K} } \ln \frac{L}{ \epsilon}+\ln \Big( \sum_{g\in \sigma} \frac{1}{ |G|}  \big[\chi_R([g]) \big]^{2K} \Big)~.
 \end{align}
Consequently, the entanglement entropy in the limit where $L$   is very large can be calculated  using (\ref{entee1}) and one obtains
 \begin{align}
 S_A = \frac{c}{ 6}  \ln  \frac{L}{ \epsilon}+ (1-\partial_K)  \ln \Big( \sum_{g\in G} \frac{1}{ |G|}  \big[\chi_R([g]) \big]^{2K} \Big)\mid_{K=1}~.
 \end{align}
This expression can be simplified  using the orthogonality formula of characters and the fact that characters of $S_N$ are real
 \begin{align}\label{charorth}
 \frac{1}{ |G|} \sum_g \chi_{R} ([g])^* \chi_{S} ([g]) = \delta_{RS}~.
 \end{align}
We arrive at the final expression for the entanglement entropy of a topological defect at the entangling surface.
\begin{align}\label{result-univ}
 S_A = \frac{c}{ 6}  \ln \frac{L}{ \epsilon} - \sum_g  \frac{1}{ |G|}  \big[\chi_R([g]) \big]^2 \ln  \Big([ \chi_R([g])]^2\Big)~.
\end{align}
Note that the expression for the g-factor is considerably more complicated than the symmetric one (\ref{gfac-sym}) which is not surprising since most of the entanglement is localized close to the boundary of the entangling surface.  

\begin{figure}[ht]
     \centering
     \begin{subfigure}[b]{0.49\textwidth}
         \centering
         \includegraphics[width=\textwidth]{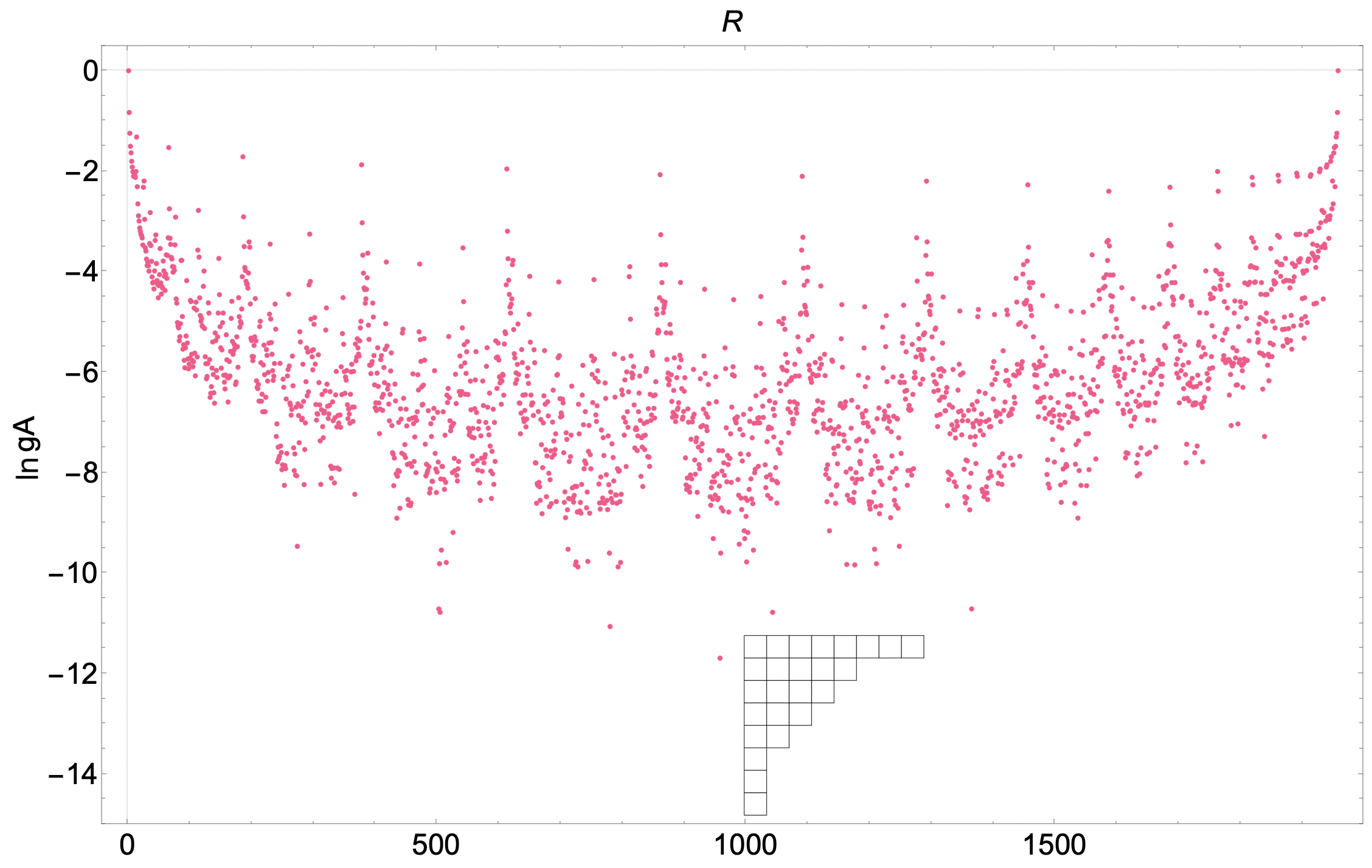}
         \caption{Defect entropy $\ln  g_{\cal A} $ for $N=25$.}
         \label{N25a}
     \end{subfigure}
     \hfill
     \begin{subfigure}[b]{0.49\textwidth}
         \centering
         \includegraphics[width=\textwidth]{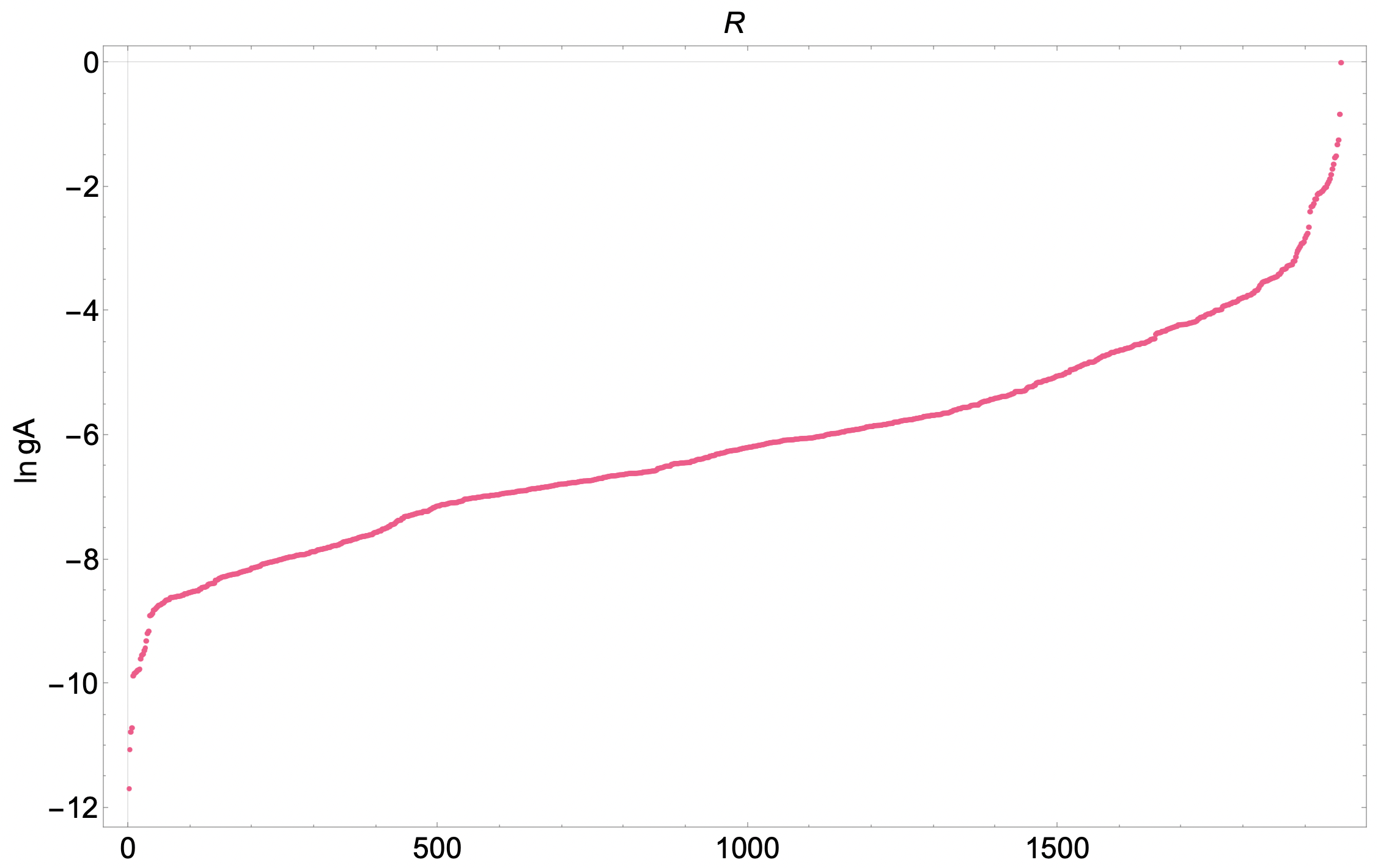}
         \caption{$\ln  g_{\cal A}$ sorted by magnitude for $N=25$}
         \label{N25b}
     \end{subfigure}
        \caption{}
        \label{figure:2}
\end{figure}

\begin{figure}[ht]
     \centering
     \begin{subfigure}[b]{0.49\textwidth}
         \centering
         \includegraphics[width=\textwidth]{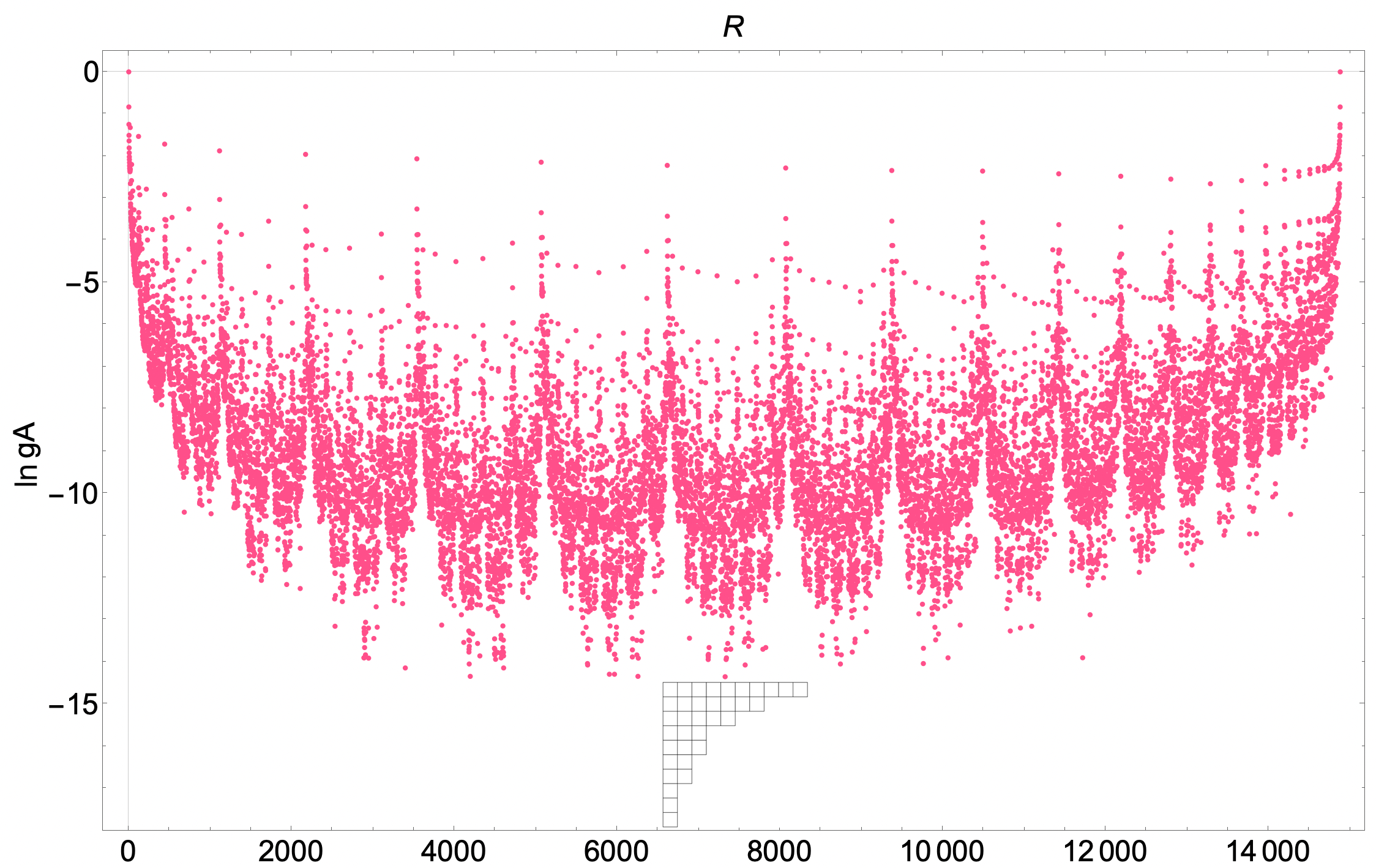}
         \caption{Defect entropy $\ln  g_{\cal A} $ for $N=35$.}
         \label{N35a}
     \end{subfigure}
     \hfill
     \begin{subfigure}[b]{0.49\textwidth}
         \centering
         \includegraphics[width=\textwidth]{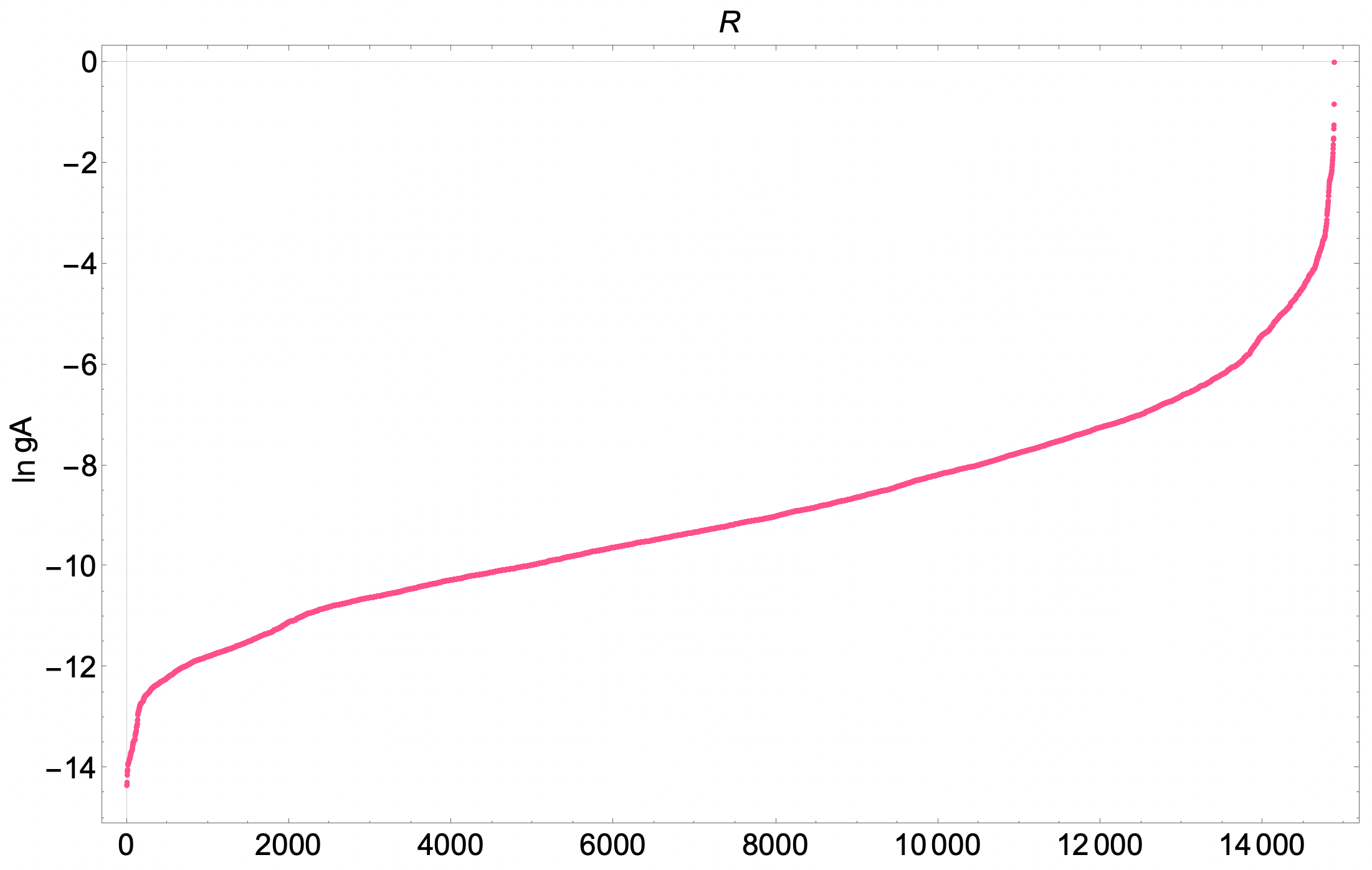}
         \caption{$\ln g_{\cal A}$ sorted by magnitude for $N=35$}
         \label{N35b}
     \end{subfigure}
        \caption{}
        \label{figure:3}
\end{figure}

We can evaluate the defect entropy from the character table of $S_N$ and the number of elements in each conjugacy class, which can be calculated using the GAP software package \cite{GAP4} for large values of $N$. In Figure \ref{figure:2} we present the results for two cases of $N=25$ for which there are $1958$ irreducible representations.  In the plot on the left the representations are dominance ordered, i.e. for two partitions $\lambda, \mu$ of $N$, one has $\lambda \trianglerighteq \mu$ if $\lambda_1+\lambda_2+\cdots +\lambda_i \geq \mu_1+\mu_2+\cdots +\mu_i$ for all $i$. In addition, we give the Young diagram associated with the representation which has the largest absolute value of $\ln g_{\cal A}$. In the plot on the right, we plot $\ln g_{\cal A}$ in order of magnitude.
 In Figure \ref{figure:3}  we present the same plots for  $N=35$, for which there are $14883$ irreducible representations.

We can recognize a $\mathbbm{Z}_2$ symmetry in these plots due to the fact that in general a new irreducible representation can be obtained from an irreducible representation by acting with the one-dimensional alternating representation. The characters of these two representations differ only by signs.  Since the characters in the formula (\ref{result-univ}) appear squared, we get the same $g_{\cal A}$ for both representations.

We can obtain a bound on the defect entropy  $\ln g_{\cal A}$ given in \eqref{result-univ} using the fact that $\chi_R(g) \leq \chi_R(1)$ for all $g\in S_N$ 
\begin{align}
   \sum_g  \frac{1}{ |G|}  \big[\chi_R([g]) \big]^2 \ln  \Big([ \chi_R([g])]^2\Big) & \leq \sum_g  \frac{1}{ |G|}  \big[\chi_R([g]) \big]^2 \ln  \Big([ \chi_R(1)]^2\Big) \nonumber\\
   &\leq  2 \ln \Big( {\rm dim}(R)\Big)~,
\end{align}
where we used character orthogonality \eqref{charorth} as well as $\chi_R(1)={\rm dim}(R)$. In \cite{vershik1981asymptotic} an asymptotic bound for  the dimension of the representation ${\rm dim}(R)$ was derived
\begin{align}
    \ln \Big( {\rm dim }(R) \Big)\leq N \ln N-N- c \sqrt{N}~,
\end{align}
where $c$ is a constant $c\sim 2.5651$. This result bounds the absolute value of the defect entropy. Numerical evaluation for $N\leq 35$ seems to indicate that this bound is not very strong and it is an interesting question whether a stronger bound exists, but we have not been able to find one in the mathematics literature.

\subsection{Non-universal defects}
For the non-universal defect, the defect symmetric entropy is given by taking the logarithm of  the g-factor for the folded boundary state (\ref{doubled-c})
\begin{align}\label{result-nonu-sym}
    g_{R,a} &= \langle 0\mid B_a^R\rangle\rangle 
    = \chi_R(1) \left( \frac{S_{a0}}{ S_{00}}\right)^N~,
\end{align}
which follows from the fact that in the untwisted sector we have $N$ copies of the RCFT defect.

For the calculation of the entanglement entropy at the defect, we have to evaluate the replicated partition function (\ref{zreplica2}) with the defect ${\cal I}$ which is obtained from unfolding the folded boundary state (\ref{doubled-b}).  Writing down the complete partition function is somewhat cumbersome due to the summation over the centralizer of $g$, but we should note that following the same strategy for calculating the entanglement as in section \ref{subsec4.1} still applies and the leading contribution is obtained from the partition function in the $h=1$ sector, before modular transformation, which gets mapped to the dominant untwisted sector after modular transformation
\begin{align}
    Z(K) = \sum_{g} \frac{1}{|G|} \big(\chi_R([g])\big)^{2K} \prod_{j=1}^N \left( \sum_i \left| \frac{S_{ai}}{ S_{0i}}\right|^{2K}  \chi_i (\frac{\tau}{ j})  \bar \chi_i (\frac{\bar \tau}{ j}) \right)^{l_j}+\cdots~.
\end{align}
The limit $\tau\to 0$ of $Z$ is evaluated by an  S-modular transformation using the transformation of the RCFT characters given in section 
\ref{subsec:nonuni}.
\begin{align}\label{znonuni1}
    \lim_{\tau \to 0}  Z(K) 
&\sim\lim_{\tau \to 0} \sum_g \frac{1}{ |G|} \big(\chi_R([g])\big)^{2K} \prod_{j=1}^N \left( \sum_{i,k,\bar k} \left| \frac{S_{ai}}{ S_{0i}}\right|^{2K}  S_{ik} \; S^*_{i \bar k} \; \chi_k (-j \frac{1}{ \tau}) \bar \chi_{\bar k} (-j \frac{1}{ \bar \tau}) \right)^{l_j}+\cdots~.
\end{align}
The dominant contribution in the $\tau\to 0$ limit in (\ref{znonuni1}) comes from the vacuum character $\chi_0$, which behaves like $q^{-\frac{c_{seed}}{ 24}} \bar q^{-\frac{c_{seed}}{  24}}$ as $q\to 0$. Using this expansion, and the fact that $\sum_i l_j =N$ for all permutations $g\in S_N,$ one obtains
\begin{align}
   \lim_{\tau \to 0}  Z(K) 
&\sim   \sum_g \frac{1}{|G|} \big(\chi_R([g])\big)^{2K} \Big( \sum_i |S_{ai}| ^{2K} |S_{0i}|^{2-2K}\Big)^N \exp\big( {\frac{c}{ 12 } \frac{1}{ K} \ln \frac{L}{ \epsilon} }\big) +\cdots~.
\end{align}
Taking the logarithm produces three terms 
 \begin{align}
  \lim_{\tau \to 0}  \ln  Z(K)&=  {\frac{c}{12 }  \frac{1}{ K} } \ln \frac{L}{ \epsilon}+\ln \Big( \sum_{g\in \sigma} \frac{1}{ |G|}  \big[\chi^R([g]) \big]^{2K}\Big) +N \ln \Big( \sum_i |S_{ai}| ^{2K} |S_{0i}|^{2-2K}\Big)+\cdots~.
 \end{align}
Consequently, the entanglement entropy in the limit where $L$   is very large can be calculated  using (\ref{entee1}) and one obtains
 \begin{align}
 S_A =  \frac{c}{ 6}  \ln \frac{L}{ \epsilon}- \sum_g  \frac{1}{ |G|}  \big[\chi_R([g]) \big]^2 \ln  \Big([ \chi_R([g])]^2\Big)- N \sum_{i} |S_{ai}|^2 \ln \left(\left| \frac{S_{ai}}{  S_{0i}}\right|^2\right)~,
 \end{align}
 where we have repeated the calculation of section \ref{subsec4.1} to obtain the second term and used the unitarity and symmetry of the modular $S$ matrix, in particular $\sum_j |S_{aj}|^2=1$  as in \cite{Brehm:2015plf,Gutperle:2015kmw}. It is interesting that for the ``maximally fractional" universal defect, the defect entropy $\ln g_{\cal A}$ is the sum of the contribution of the universal defect  (\ref{result-univ}) and $N$ times the contribution of the RCFT topological defect \cite{Brehm:2015plf,Gutperle:2015kmw}.  This together with the result for the symmetric entanglement entropy (\ref{result-nonu-sym}) indicates that the maximally fractional defect can be in some sense viewed as a product of the RCFT and symmetric orbifold defect.
It would be interesting to see whether more general constructions,
maybe along the lines of constructions for D-branes in symmetric orbifolds presented in \cite{Belin:2021nck}, are possible and lead to a more complicated interplay between the RCFT defect and the universal defect.

\section{Discussion}
\label{sec:Discussion}

In this note we have calculated the sub-leading constant term (the g-factor or defect entropy) of the entanglement entropy of a single interval in the presence of topological defects for symmetric orbifold CFTs. For the symmetrically placed topological defect (or away from the boundary of the entangling region), this is simply the quantum dimension of the defect. For the defect placed at the boundary, the behavior of $g_{\cal A}$ is more complicated, with a formula that is reminiscent of a classical Shannon entropy with the characters of $S_N$  and the modular $S$ matrix taking the role of a probability distribution.  We used numerical methods to evaluate the defect entropy for $S_N$ with $N\leq 35$. It would be interesting, in particular for applications to holography,  to investigate whether one can take the $N\to \infty $ limit of the characters using the machinery of \cite{okounkov1998representations, Borodin}. 

We considered non-universal defects that use both the universal $Rep(S_N)$ defects and topological defects in the seed CFT. It would be interesting to consider topological defects in seed CFTs relevant for AdS/CFT, namely supersymmetric sigma-models in $T^4$ or $K_3$. The behavior of the entanglement entropy for non-universal defects factorizes, it would also be interesting to find defects that have a more complicated interplay between $S_N$ and the seed CFT, although the construction of defects which satisfy the Cardy-Petkova-Zuber \cite{Cardy:1989ir, Petkova:2000ip} consistency condition is a challenge at this point.

The behavior of the entanglement entropy with topological defects is particularly simple in the sense that the logarithmic divergence depends only on the central charge of the orbifold CFT. For conformal defects, the central charge is replaced by an effective central charge which depends on the details of the defect in a complicated fashion \cite{Sakai:2008tt}. It would be interesting whether there are simple conformal defects sharing properties of our universal topological defects for which the effective central charge can be computed. This would be particularly interesting in the light of AdS/CFT since these defects should correspond to branes in the bulk \cite{Gaberdiel:2021kkp,Belin:2021nck} with finite tension.

We leave these interesting questions for future work.

\acknowledgments
The work of M.G. was supported, in part, by the National Science Foundation under grant PHY-2209700. 
M.G., Y.L and D.R. are grateful to the Bhaumik Institute for support. K.R. is supported by the Simons Collaboration on Global Categorical Symmetries and also by the NSF grant PHY-2112699.   M.G. is grateful to the Harvard Center for the Fundamental Laws of Nature for hospitality while this paper was finalized.

\newpage

\providecommand{\href}[2]{#2}\begingroup\raggedright\endgroup

\end{document}